\documentstyle[12pt]{article}
\advance\hoffset by -7mm  
\makeatletter
\@addtoreset{equation}{section}
\makeatother

\renewcommand{\title}[1]{\null\vspace{25mm}

\noindent{\Large{\bf #1}}\vspace{10mm}

\noindent {\large By }}

\renewcommand{\abstract}[1]{\vspace{19mm}

\noindent{\small{\em Abstract.} #1}\vspace{2mm}

} 

\begin{document}

\def\lin{\bigskip\centerline{\vrule width 5cm height0.4pt}\bigskip}
\def\d{\partial}
\def\dh{\mathop{\vphantom{\odot}\hbox{$\partial$}}}
\def\dl{\dh^\leftrightarrow}
\def\sqr#1#2{{\vcenter{\vbox{\hrule height.#2pt\hbox{\vrule width.#2pt 
height#1pt \kern#1pt \vrule width.#2pt}\hrule height.#2pt}}}}
\def\w{\mathchoice\sqr45\sqr45\sqr{2.1}3\sqr{1.5}3\,} 
\def\fii{\varphi}
\def\eps{\varepsilon}
\def\hq{\hbar}
\def\lo{\cal L_+^\uparrow}
\def\psq{{\overline{\psi}}}
\def\pp{\psi ^\prime}
\def\ppq{\overline{\psi ^\prime}}
\def\sp{\vec\sigma\cdot\vec p}
\def\pdh{(2\pi )^{-3/2}}
\def\ps{\hbox{\rm\rlap/p}}
\def\ec{{e\over c}}
\def\=d{\,{\buildrel\rm def\over =}\,}
\def\iix{\int d^3x\,}
\def\iip{\int d^3p}
\def\inx{\int d^3x_1\ldots d^3x_n}
\def\H{{\cal H}}
\def\F{{\cal F}}
\def\N{\hbox{\bbf N}}
\def\A{\hbox{\bbf A}}
\def\xn{\vec x_1,\ldots ,\vec x_n\,}
\def\vxp{\vec x\,'}
\def\V{\hbox{\bbf V}}
\def\S{\hbox{\bbf S}}
\def\U{{\hbox{\bbf U}}}
\def\HH{\hbox{\bbf H}}
\def\Q{\hbox{\bbf Q}}
\def\i3p{\p32\int d^3p}
\def\psm{\psi ^{(-)}}
\def\psp{\psi ^{(+)}}
\def\px{\vec p\cdot\vec x}
\def\pqp{\overline{\psi}^{(+)}}
\def\pqm{\overline{\psi}^{(-)}}
\def\vq{\overline v}
\def\uq{\overline u}
\def\iep{\int {d^3p\over 2E}}
\def\ipd{\int d^4p\,\delta (p^2-m^2)}
\def\ds{\hbox{\rlap/$\partial$}}
\def\pe{\sqrt{\vec p^2+m^2}}
\def\dsx{\hbox{\rlap/$\partial$}_x}
\def\itn{\int\limits_{-\infty}^{+\infty}dt_1\int\limits_{-\infty}^{t_1}
dt_2\cdots\int\limits_{-\infty}^{t_{n-1}}dt_n}
\def\ipn{\int d^3p_1\cdots\int d^3p_{n-1}\, }
\def\As{A\hbox to 1pt{\hss /}}
\def\np4{\int d^4p_1\cdots d^4p_{n-1}\, }
\def\Sr{S^{\rm ret}}
\def\gp{\vec\gamma\cdot\vec p}
\def\te{\vartheta}
\def\tr{{\rm tr}\, }
\def\Sa{S^{\rm av}}
\def\qs{\hbox{\rlap/q}}
\def\supp{{\rm supp}\, }
\def\Tr{{\rm Tr}\, }
\def\Im{{\rm Im}\, }
\def\sgn{{\rm sgn}\, }
\def\cau{{1\over 2\pi i}}
\def\P{\,{\rm P}}
\def\Re{{\rm Re}\, }
\def\iinf{\int\limits_{-\infty}^{+\infty}}
\def\kx{\vec k\cdot\vec x}
\def\io{\int{d^3k\over\sqrt{2\omega}}\,}
\def\nx4{\int d^4x_1\ldots d^4x_n\, }
\def\xnn{x_1,\ldots ,x_n}
\def\xnm{x_1,\ldots ,x_{n-1},x_n}
\def\Dr{D^{\rm ret}}
\def\Da{D^{\rm av}}
\def\kon#1#2{\vbox{\halign{##&&##\cr
\lower4pt\hbox{$\scriptscriptstyle\vert$}\hrulefill &
\hrulefill\lower4pt\hbox{$\scriptscriptstyle\vert$}\cr $#1$&
$#2$\cr}}}
\def\lra{\longleftrightarrow}
\def\konv#1#2#3{\hbox{\vrule height12pt depth-1pt}
\vbox{\hrule height12pt width#1cm depth-11.6pt}
\hbox{\vrule height6.5pt depth-0.5pt}
\vbox{\hrule height11pt width#2cm depth-10.6pt\kern5pt
      \hrule height6.5pt width#2cm depth-6.1pt}
\hbox{\vrule height12pt depth-1pt}
\vbox{\hrule height6.5pt width#3cm depth-6.1pt}
\hbox{\vrule height6.5pt depth-0.5pt}}
\def\konu#1#2#3{\hbox{\vrule height12pt depth-1pt}
\vbox{\hrule height1pt width#1cm depth-0.6pt}
\hbox{\vrule height12pt depth-6.5pt}
\vbox{\hrule height6pt width#2cm depth-5.6pt\kern5pt
      \hrule height1pt width#2cm depth-0.6pt}
\hbox{\vrule height12pt depth-6.5pt}
\vbox{\hrule height1pt width#3cm depth-0.6pt}
\hbox{\vrule height12pt depth-1pt}}
\def\es{\hbox{\rlap/$\varepsilon$}}
\def\ks{\hbox{\rlap/k}}
\def\konw#1#2#3{\hbox{\vrule height12pt depth-1pt}
\vbox{\hrule height12pt width#1cm depth-11.6pt}
\hbox{\vrule height6.5pt depth-0.5pt}
\vbox{\hrule height12pt width#2cm depth-11.6pt \kern5pt
      \hrule height6.5pt width#2cm depth-6.1pt}
\hbox{\vrule height6.5pt depth-0.5pt}
\vbox{\hrule height12pt width#3cm depth-11.6pt}
\hbox{\vrule height12pt depth-1pt}}
\def\grad{{\rm grad}\, }
\def\diw{{\rm div}\, }
\def\eh{{\scriptstyle{1\over 2}}}
\def\intt{\int dt_1\ldots dt_n\,}
\def\tnn{t_1,\ldots ,t_n}
\def\dett{{\rm det}\,}
\def\lap{\bigtriangleup\,}
\def\HHe{\hbox{\bex H}}
\font\bex=cmbx7
\def\Wiout{W_{{\rm\scriptstyle in}\atop{\rm\scriptstyle out}}}
\def\Win{W_{\rm in}}
\def\Wout{W_{\rm out}}
\def\poi{\cal P _+^\uparrow}
\def\gu{\underline{g}} 
\def\hu{\underline{h}}
\def\i{{\rm int}}
\def\su{\sum_{n=1}^\infty}
\def\suu{\sum_{n=0}^\infty}
\def\c{{\rm cl}}
\def\e{{\rm ext}}
\def\snn{\su {1\over n!}\nx4}
\def\r{{\rm ret}}
\def\a{{\rm av}}
\def\ra{{{\rm\scriptstyle ret}\atop{\rm\scriptstyle av}}}
\def\dsu{\mathop{\bigoplus}}            
\def\m3{{\mu_1\mu_2\mu_3}}
\def\itt{\int\limits_{t_1}^{t_2}}
\def\gs{\,{\scriptstyle{>\atop <}}\,}
\def\ga{\,{\scriptstyle{>\atop\sim}\,}}
\def\Seh{{\cal S}^\eh_\eh}
\def\SS{{\cal S}({\bf R^4})}
\def\co{{\rm Com}}
\def\ml{{m_1\ldots m_l}}
\def\ik{{i_1\ldots i_k}}
\def\jt{{j_1\ldots j_t}}
\def\is{{i_{k+1}\ldots i_s}}
\def\js{{j_{t+1}\ldots j_s}}
\def\xnr{x_1-x_n,\ldots x_{n-1}-x_n}
\def\qm{q_1,\ldots q_{m-1}}
\def\pn{p_1,\ldots p_{n-1}}
\def\xll{x_l-x_{l+1}}
\def\xmm{x_1,\ldots x_{n-1}}
\def\ph{{\rm phys}}
\def\nab{\bigtriangledown}
\def\p{{(+)}}
\def\ul{\underline}
\def\tu{\tilde u}

\pagestyle{empty}
\vbox to 3,0cm{ITP-SB-96-50, to appear in HPA }
\centerline{\Large\bf Higgs-free Massive Nonabelian  } 
\vskip 0.3cm
\centerline{\Large\bf Gauge Theories \footnote{Work supported by the Swiss National Science Foundation}} 
\vskip1,5cm
\centerline{\large\bf Tobias Hurth \footnote{ Email: hurth@insti.physics.sunysb.edu}} 
\vskip 0.5cm
\centerline{\large\it
Institute for Theoretical Physics, SUNY at Stony Brook}
\centerline{\large\it Stony Brook, New York 11794-3840}
\vskip 0,1cm
\vskip 1,0cm

{\bf Abstract.} - We analyze nonabelian massive Higgs-free theories 
in the causal Epstein-Glaser approach. 
Recently, there has been renewed interest 
in these models. In particular we consider the well-known 
Curci-Ferrari model and the nonabelian St\"uckelberg models. 
We explicitly show the reason why  the considered models fail 
to be unitary. In our approach only the asymptotic (linear)
BRS-symmetry has to be considered. \\

{\bf PACS.} 11.10 - Field theory, 12.10-Unified field theories and models.
\vskip 3.5cm
$^*)$ Emailaddress: hurth@insti.physics.sunysb.edu\\

\newpage




The discription of massive gauge bosons favoured today is the 
Higgs-Kibble mechanism: It is the only mechanism of mass generation
 known so far which leads to a normalizable and unitary theory of 
massive nonabelian gauge bosons.  One introduces new 
spin-zero-particles (Higgs-fields) with unknown mass and 
couplings into the theory for which there are no experimental 
evidence so far. But, for instance, the measured ratio of the W- 
and Z-boson masses for example is at least a phenomenological 
indication that these masses are generated by spontaneously 
symmetry breaking. In recent years the classical Higgs field
 has become available for a geometrical interpretation as 
generalized connection in the context of non-commutative 
geometry [1], which makes the models more attractive.
\\
However, the nondiscovery of Higgs bosons and the 
well-known shortcomings in this approach 
(for example the hierarchy problem) lead to 
continued attempts to construct alternative 
massive nonabelian gauge theories (see [2] for 
a review). Two prominent approaches are the 
Curci-Ferrari model [3] and the nonabelian 
generalization of the massive abelian St\"uckelberg 
gauge theory [4,5]. The findings suggest that the 
properties of perturbative normalizability and of 
physical unitarity are mutually exclusive. Moreover, 
gauge invariance and gauge independence are described 
as necessary but not sufficient conditions for physical 
	unitarity, i.e. for decoupling of unphysical 
degrees of freedom in the theory [2]. Nevertheless, 
there has been renewed interest in these models. In fact, 
Periwal [6] has proposed a nonperturbative condition on a 1PI 
distributions for physical unitarity which fixes the gauge 
parameter $\xi$ in a nonlinearly gauged Curci-Ferrari model. 
But the nonunitarity of this Curci-Ferrari model, for arbitrary 
values of the parameters of the theory, was  quite recently 
reassured by improving Ojima's proof [7] of this statement [8].\\
\\
Because of the frequent questioning of the non-unitarity results
we want to give a brief reanalysis of these models using the 
Epstein-Glaser methods in this letter. \\ 

The causal Epstein-Glaser formalism [9,see also 10] represents a general framework for perturbative quantum field theory.
The method allows for a clear and simplified analysis of these models which
 accurately spells out the reasons for the absence of unitarity in these models. The analysis is simplified by the fact that only the asymptotic (linear) part
of the BRS-transformations is relevant in this approach.
\\
In the causal approach the technical details 
concerning the well-known 
UV- and IR-problem in quantum field theory are separated 
and reduced to
mathematically well-defined problems, namely the 
causal splitting  and the adiabatic switching of 
operator-valued distributions.\\
 The $S$-matrix is directly constructed in the well-defined 
Fock space of free asymptotic fields in the form of a 
formal power series
$$S(g)=1+\sum_{n=1}^{\infty}{1\over n!}\int d^4x_1...
d^4x_n\,T_n(x_1,...,x_n)g(x_1)...g(x_n),\eqno(1)$$
where $g(x)$ is a tempered test function which switches 
the interaction. 
Only well-defined free field operators occur in the 
whole construction.
The central objects are the $n$-point distributions $T_n$. 
They may be viewed as mathematically well-defined 
time-ordered products.
The defining equations of the theory in the causal 
formalism are the fundamental (anti-) commutation 
relations of the free field operators, their dynamical 
equations and the specific coupling of the theory $T_{n=1}$. 
The $n$-point distributions $T_n$ in (1) are then constructed 
inductively from the given first order $T_{n=1}$. Epstein and 
Glaser present an explicit inductive construction of the 
general perturbation series in the sense of (1) which is 
compatible with causality and Poincare invariance.\\
The causal formalism allows for a comprehensive discussion  
of massless Yang-Mills theories in four (3+1) dimensional 
space time (see [11,12,13] for details). 
It was shown that the whole analysis of nonabelian gauge 
symmetry can be done in the well-defined Fock space of
free asymptotic fields. The LSZ-formalism is not necessary then. 
Nonabelian gauge invariance is introduced  by a linear operator 
condition in every order of perturbation theory separately:
$$[Q,T_n(x_1,...,x_n)]= d_Q  T_n(x_1,.....,x_n)= 
i\sum_{l=1}^n\d_\mu^{x_l}T^
\mu_{n/l}(x_1,...,x_n)\equiv div. \eqno(2)$$
where the charge $Q$ is the generator of  the linear (abelian!)
 BRS transformations of the free asymptotic field operators which 
defines an antiderivation $d_Q$ in the  algebra, generated by the 
fundamental field operators. The  $T_{n/l}^\nu (x_1, \ldots, x_n)$ 
are $n$-point distributions of an extended theory which also can 
be inductively constructed in the causal formalism. They serve 
for an explicit representation of
 the commutator $[Q,T_n(x_1,...,x_n)]$ as a divergence  in the sense of 
vector analysis.\\ Physical unitarity, i.e.  decoupling of the unphysical 
degrees of freedom, is shown as a direct consequence of the  linear 
operator
gauge invariance condition (2) and of the nilpotency of the
charge $Q$.
Perturbatively, physical unitarity means
$$\tilde {T}_n^{P_\bot} = P_\bot T_n^+ P_\bot + div 
\quad \forall n \eqno(3)$$
where $div$ denotes distributions of divergence form as in 
the condition of gauge invariance (2), $P_\bot$ is the 
projection operator on the physical subspace, $+$ denotes the 
hermitean conjugation with regard to the Hilbert scalar 
product of the Fock space. The $\tilde {T}_n^{P_\bot}$ are 
the $n$-point distributions of the inverse 
$(P_\bot S(g) P_\bot)^{-1}$-matrix
restricted to the physical subspace:
$$\big{(} P_\bot S(g) P_\bot \big{)}^{-1} = \sum_{n} 
\frac{1}{n!} \int d^4x_1 \ldots \int d^4x_n \quad 
\tilde{T}_n^{P_\bot} (x_1, \ldots, x_n) g(x_1)\ldots g(x_n). $$
The $n$-point distributions $\tilde{T}_n^{P_\bot}$ are computed 
by formal inversion of (1). They are equal to the 
following sum over subsets of\\ $X$ $= \{x_1, \ldots, x_n\}$
$$\tilde{T}_n^{P_\bot}(X) = \sum_{r=1}^{n} (-)^r \sum_{Pr}
 P_\bot T_{n_1}(X_1) P_\bot \ldots P_\bot T_{n_r} (X_r) P_\bot. $$
The perturbative statement (3) implies the 
following statement about  a formal power series:
$$ \quad (S_\bot)^{-1}(g) =S_\bot^+(g) + div(g) \qquad  
\qquad S_\bot = P_\bot SP_\bot \eqno(4)$$
Finally, normalizability of the theory means in the 
Epstein-Glaser approach that  the number of the  
finite constants to be fixed by physical conditions 
stays the same in all orders  of perturbation theory. This property is based on scaling properties of the theory only.  
The following conditions are shown to be sufficient for
 this property :\\
 (a) The specific coupling $T_{n=1}$ of the theory has 
maximal mass dimension four and\\
 (b) the singular order of the fundamental (anti-) 
commutator distributions  of the free asymptotic fields
 are smaller than zero.\\
Note that in this context normalizability does not necessarily mean 
that the theory can be normalized in a gauge invariant way, a more far
 reaching quality generally referred as renormalizability.\\
\\
Considering genuine massive nonabelian gauge theories, normalizability is established {\it per definitionem} in the theory by
 suitable choice of the defining equations. For example , we choose 
the following commutator relations of the asymptotic field 
operators and their corresponding equation of motion.
The massive gauge potentials in a general linear $\xi-$gauge, transforming according to the adjoint 
representation of $SU(N)$, 
satisfy 
$$ (\Box + m^2) A_\mu^a (x) - (\frac{\xi -1}{\xi}) \d_\mu 
(\d^\nu A_\nu) = 0. \eqno(5)$$
$$\left[ A_\mu^a (x), A_\nu^b (y) \right]_{-} = i \delta_{ab} 
( g_{\mu\nu} +
\frac{\d_\mu  \d_\nu}{m^2}) D_m (x-y) -  i \delta_{ab} 
\frac{\d_\mu  \d_\nu}{m^2} 
D_M (x-y) \eqno(6)$$
$D_m$ denotes the Pauli-Jordan commutation distribution with 
mass $m$.\\
The masses $m$ and $M$ are related: $$ M^2 = m^2 \xi \eqno(7)$$
The right side of (6) represents the general ansatz compatible
with normalizability, Poincare invariance, field equation (5) 
and causality.\\
The ghost fields may fulfill (in a general $\xi$-gauge):
$$\left\{ u_a (x), \tilde{u}_b (y) \right\}_{+} = -i \delta_{ab} 
D_M (x-y) \eqno (8) $$
$$(\Box + M^2) u_a (x), \quad (\Box + M^2) 
\tilde{u}_a (x) = 0.  \eqno (9)$$ 
Note that this relation between the masses of the gauge 
bosons and the ghosts
is already suggested by the most general gauge invariant quadratic 
terms in the specific coupling in the massless theory 
(see Lemma 3.1 and 4.1 in [13]) This relation is uniquely 
fixed by gauge invariance.\\
It is well-known that the gauge boson field can be splitted 
into the Proca component (representing the three physical 
transverse components) and 
the unphysical part:
$$ A_\mu^a = A_\mu^{a,phys} - \frac{1}{m^2 \xi} \d_\mu 
(\d^\nu A_\nu^a) \eqno(10)$$    
$$\left[ A_\mu^{a,phys} (x), A_\nu^{b,phys} (y) \right]_{-} = 
i \delta_{ab} ( g_{\mu\nu} +
\frac{\d_\mu  \d_\nu}{m^2}) D_m (x-y) \eqno(11)$$
\\
Having defined the Fock space of free asymptotic 
fields  by the first two defining
equations (5) and (6) we can further pursue the 
standard procedure in the causal formalism [13]:
We choose a reasonable gauge invariance condition 
and then construct the most general gauge invariant 
specific coupling, the third defining equation of the 
theory in the causal formalism.\\
\\
 As has been firstly noticed by Curci and Ferrari 
[2, see also 7], a direct taking over of the formula of the 
generator $Q_{CF}$ from the massless case  in a general 
$\xi$-gauge [13]
$$Q_{CF} = \int    \frac{\partial_\nu A^\nu}{\xi}
 \stackrel{\leftrightarrow}{\partial}_0 u d^3 \bar{x}  
\eqno (12)$$
 
leads to a missing nilpotency of $Q_{CF}$ in the 
massive case. One easily checks that $Q_{CF}^2$ is 
proportional to $m^2$, because of 
$\left[ \d^\mu A_\mu^a (x), \d^\nu A_\nu^b (y) 
\right]_{-} \neq 0 $.\\ 
But our analysis of the massless case shows that
the nilpotency of $Q_{CF}$ is the crucial input to 
determine unitarity in the physical subspace (3) as a 
direct consequence of the operator gauge invariance 
condition (2). So this gauge invariance
condition does not seem to be very useful. Nevertheless, 
at the end of this letter we will come back to these 
specific couplings, which are gauge invariant in respect 
to the charge $Q_{CF}$ in (12).\\
\\
St\"uckelberg's idea, generalized to the nonabelian 
case [4,5], is to introduce an additional scalar 
field $\zeta^a (x)$ transforming also according to 
the adjoint representation
$$\left[ \zeta_a (x), \zeta_b (y) \right] = -i 
\delta_{ab} D_M (x-y) \qquad (\Box + M^2) 
\zeta_a (x) = 0  \eqno (13)$$

Note that we have chosen the mass $M$ of
 the unphysical component of the 
gauge bosons and the ghosts.
Now we introduce also generalized BRS transformations 
of the free asymptotic fields which involve this new field 
$\zeta(x)$ [5].\\
The corresponding generator $Q^s$ of these St\"uckelberg 
gauge transformations in the Fock space of asymptotic 
field is (see formula 3.31 in [5]; note that we can leave 
out the Z-factors because we directly work in the 
well-defined Fock space of free asymptotic fields and 
does not use the LSZ-formalism, moreover we have 
generalized the formula 3.31 in [5] to a general 
linear $\xi$-gauge):

$$Q_s = \int \eta^a (x) \stackrel{\leftrightarrow}{\partial}_0 
u_a (x) d^3 \bar{x}$$
$$\mbox{with} \quad \eta^a (x) := 
\frac{\partial_\mu A_a^\mu (x)}{\xi} + 
m \zeta^a (x)  \eqno (14)$$

As one easily verify, we have $Q_s^2 = 0$
 because of $\left[ \eta (x), \eta (y) \right]_{-} = 0 $ 
and arrive at a well-defined anti-derivation $d_{Q^s}$ 
in the graded algebra of fields: The gradation is 
introduced by the ghost charge  
$$Q_c:=i \int d^3x :(\tilde u \stackrel{\leftrightarrow}{\d}_0 u): . $$ 

The anti-derivation $d_{Q_s}$ in the graded algebra 
is then given by 
$$d_{Q^s} \hat A:= Q_s \hat A-(e^{i\pi Q_c} \hat A 
e^{-i\pi Q_c}) Q_s
$$
with 
$$d_{Q^s} A_a^\mu = i \partial^\mu u_a, 
\quad d_{Q^s} u_a = 0,\quad  d_{Q^s} \tilde{u}_a = 
-i \eta_a,$$ $$ \quad d_{Q^s} \zeta_a = im u_a, 
\quad d_{Q^s} \d^\mu A_\mu^a = - i M^2 u_a, 
\quad d_{Q^s} \eta_a = 0. \eqno(15)$$
 According to the standard procedure  in the 
causal formalism [13] we construct the most general 
gauge invariant specific coupling with respect to 
this antiderivation:\\
{\bf Lemma:} The most general gauge invariant 
coupling $T_1^g$, {\bf (A)} $\quad  d_{Q^s} 
T_1^g=div$, which is also invariant under 
the special Lorentz group {\bf $L_+^\uparrow$} 
{\bf (B)} and  under the structure group $G=SU(N)$ 
{\bf (C)}, which  has ghost number zero - $G(T_1^g)=0$ - 
{\bf (D)}  and has maximal mass dimension 4  {\bf (E)} 
and is invariant under the discrete symmetry 
transformations {\bf  (F)} can be written as
$$T_1^g = -igf_{a'b'c'} : A_\kappa^{a'} A_\lambda^{b'} 
\partial_\kappa A_\lambda^{c'} :  - \frac{1}{2} igf_{a'b'c'}
 : A_\kappa^{a'} u_{b'} \partial_\kappa \tilde{u}_{c'} :+
 \eqno(16.a.b.)$$  $$+ \frac{1}{2} igf_{a'b'c'} : 
A_\kappa^{a'} \partial^\kappa u_{b'} \tilde{u}_{c'} : 
+ \frac{1}{2} igf_{a'b'c'} : A_\kappa^{a'} \zeta_{b'} 
\partial_\kappa \zeta_{c'} : +\eqno(16.c.d.)$$  $$+ 
\alpha \quad \partial_\kappa \left[ igf_{a'b'c'} : 
A_{a'}^\kappa u_{b'} \tilde{u}_{c'} : \right] + 
\beta \quad d_{Q^s} \left[ gf_{a'b'c'} : u_{a'} \tilde{u}_{b'} 
\tilde{u}_{c'} : \right] \eqno (16.e.f.)$$
The explicit representation of $d_{Q^s} T_1^g$ as a 
divergence is given by 
$$d_{Q^s} T_1^g = i \partial_\mu T_{1,g}^\mu + i 
\gamma \quad \partial_\mu B_{1,g}^\mu \eqno (17)$$
$$T_{1,g}^\mu = - igf_{abc} : u_a A_\nu^b ( \partial_
\mu A_\nu^c - \partial_\nu A_\mu^c) : - i \frac{1}{2} 
gf_{abc} : u_a u_b \partial^\mu \tilde{u}_c :+ \eqno (18.a.b.c.)$$ 
$$+ i \frac{1}{2} gf_{abc} : u_a \partial^\mu u_b \tilde{u}_c :
- i \frac{1}{2} gf_{abc} : u_a A_b^\mu \partial_\nu A_c^\nu :+ 
\eqno(18.d.e.)$$ $$+\frac{1}{2} igf_{abc} : u_a \zeta_b \partial^\mu 
\zeta_c : + \eqno(18.f.)$$ $$
+ i \alpha \left[ gf_{abc} : \partial_\mu u_a u_b \tilde{u}_c :+ 
\frac{1}{\xi} gf_{abc} : A_\mu^a u_b \partial_\kappa A_c^\kappa:  
+ gf_{abc} : A_\mu^a u_b m \zeta_c: \right] \eqno (18.g.h.i.) $$
$$  B_{1,g}^\mu =  \partial_\nu \left\{ gf_{abc} u_a A_b^\mu A_c^\nu : 
\right\}  \eqno(18.j.)$$
$\alpha,\beta,\gamma$ are free constants. \\
The {\bf proof} of this statement is straightforward and analogous 
to the one in the massless case (see Appendix A of [13]).
\\
Note that all Lorentz invariant {\bf (B)}, G-invariant 
{\bf (C)} terms with ghost number zero {\bf (D)} and with 
{\bf four} normalordered operators which would be compatible 
with normalizability {\bf (E)} and are invariant under 
the discrete symmetry transformations {\bf (F)} are ruled out 
by the gauge invariance condition {\bf (A)}. \\
Moreover we left out the quadratic terms compatible with
the conditions {\bf (A)-(F)} in formula (16) because in the 
causal formalism the information about such quadratic terms 
is already contained in the fundamental (anti-)commutation 
relations and the dynamical equations for the operators.\\
In addition, $T_{1}^g$ in (16) is also anti-gauge invariant
in respect to the anti-charge $$Q_s = \int \eta^a (x)
\stackrel{\leftrightarrow}{\partial}_0 \tilde{u}_a (x)
d^3 \bar{x} \quad \mbox{with} \quad \bar Q_s^2=0 ) :\quad $$
$$  [ \bar Q_s , T_1^g] = div \eqno(19) $$
We have thus defined a manifestly normalizable theory which 
is gauge invariant to first order of perturbation theory and 
respects certain further symmetry 
conditions. We now have to examine if 
one can prove a corresponding condition of gauge 
invariance to all orders of perturbation theory inductively
$$d_{Q^s} T_n = div \eqno(20)$$
Before studying this explicitly, we 
should emphasize that the unitarity of the 
$S$-matrix in the physical subspace would be a direct 
consequence of such a condition (20) - analogously 
to the massless case. The three physical components of the massive 
gauge boson would decouple from all other fields.
The inductive proof of this statement is completely analogous to 
the one in the massless case (see chapter 5 of [12]): Again, 
the crucial point is the fact that the 
physical subspace Ker$N$ of the Fock space has 
the following representation (see formula 3.34 of [5])
$$\mbox{Ker} N = \mbox{Ker} Q_s / \mbox{Range} Q_s \eqno (21)$$
where $N$ is the number operator of the unphysical 
particles only ( excluding the three physical 
components of the massive gauge boson).
\\
Knowing this fact (21), we could repeat the 
proof of unitarity in the physical subspace 
worked out in the massless case without any changes 
(see [12], Chapter 7), provided equation (20) holds! 
Another proof of this implication can be found in [14].
From the perspective of the causal formalism  the operator 
gauge invariance condition (20)  in the St\"uckelberg model 
is sufficient for the unitarity of the $S$-matrix in 
the physical subspace, that means that the 
perturbative condition (3) holds in every order of perturbation 
theory. However, we now show that the operator gauge invariance
 condition (20) is already violated in the tree contribution 
at second order of  perturbation theory:
\\
We prove that there is no normalization of $T_{n=2} 
\big|_{tree}$ which is gauge invariant, i.e. $d_{Q_s} T_{n=2} 
\big|_{tree} = div$. The latter statement is equivalent 
to the insolvability of the corresponding 
anomaly equation (see [13])
$$d_{Q_s} N - 2 A \stackrel{!}{=} div \eqno (22)$$
where $N$ represents free local normalization terms 
in $T_{n=2} \big|_{tree,4}$ and $A$ represents 
the local anomaly terms which arise in the natural 
splitting in second order of perturbation theory in 
order to construct $T_{n=2} \big|_{tree,4}$ [15]:\\
According to Epstein-Glaser method one has to construct the causal
commutator $$D_{n=2} (x,y) =\bigl[ 
T_1^g(x),T_1^g(y) \bigr],\eqno(23)$$   
in order to arrive at $T_{n=2}$. One verifies that gauge 
invariance of the causal commutator $D_{n=2}$ is a direct 
consequence of gauge invariance in first order (17):
$$d_{Q_s} D_{n=2}(x,y)=[Q_s,\,[T_1^g(x),\,T_1^g(y)]]=$$
$$=i \d_\nu^x([T_{1,g}^\nu(x),T_1^g(y)])
+i \d_\nu^y([T_1^g(x),T_{1/g}^\nu(y)])\eqno(24)$$
The question is whether the same (divergence form) is true
 for the commutator $[Q_s,\,R_2(x,y)]$ obtained by causal splitting 
of $[Q_s,\,D_2(x,y)]$ into a retarded and a advanced 
distribution. There is only one mechanism to spoil gauge 
invariance in the tree contribution [14]. The unique splitting 
solution of the Pauli-Jordan distribution   
$$D(x-y)=\Dr (x-y)-\Da (x-y)\eqno(25)$$  lead to a 
local term in the gauge invariance condition because
instead of $$( \Box + m^2 ) D_m(x-y)=0\eqno(26)$$ 
we have after natural splitting
$$ ( \Box + m^2 ) D_m^{ret} (x-y)=\delta(x-y).\eqno(27)$$
Consequently, the procedure is straightforward: We have to 
pick up all local terms $A$ arising in the natural splitting of 
the causal distribution $d_{Q^s} D_{n=2}$ and compare 
them with the free normalization terms $N$ in $T_{n=2}$
in order to find a solution of the anomaly equation (22).\\ 
For this purpose,  we focus on the local operator 
terms proportional to $:u A_\nu \partial_\nu \zeta \zeta:$.
Because of the constraint of normalizability 
which implies that the maximal mass dimension of the 
anomaly must be 5, there are exactly two independent 
operator terms in this sector. Thus the most general 
anomaly term arising in the natural splitting can be 
written as
$$A \big|_{u\partial_\nu 
\zeta A^\nu \zeta}  = \alpha_1 : u_a \zeta_b A_\nu^{a'} 
\partial^\nu \zeta_{b'} : f_{abc} f_{a'b'c} \delta +$$ 
$$ \alpha_2 : u_a \partial^\nu 
\zeta_b A_\nu^{a'} \zeta_{b'} : f_{abc} f_{a'b'c} 
\delta . \eqno(28)$$
Since the operator $\partial_\nu \zeta$ cannot be 
represented by a variation of any fundamental 
field (see 15), the term $d_Q N$ cannot contribute 
to the  sector $:u  A_\nu\partial_\nu \zeta  \zeta:$.
As a consequence, operator gauge invariance implies 
$$A  \big|_{u\partial_\nu 
\zeta A^\nu \zeta} 
 \stackrel{!}{=} div \big|_{u\partial_\nu 
\zeta A^\nu \zeta} \eqno (29)$$
instead of (22). The subscript on the right hand side of (29) of course means
that one has to keep only these terms of the 
total derivative $div$ which contributes to the specified sector. There is only one divergence term contributing to this  sector, namely 

$$ A \big|_{u\partial_\nu 
\zeta A^\nu \zeta}  = \partial_\nu \left[ : u_a \zeta_b  A_{a'}^\nu \zeta_{b'} : 
f_{abc} f_{a'b'c} \delta  \right] \big|_{u\partial_\nu 
\zeta A^\nu \zeta}
 \eqno (30)$$
Because of (30), equation (29) implies that $\alpha_1 = 
\alpha_2$. In the following, we explicitly calculate 
$\alpha_1$ and $\alpha_2$, and check this necessary 
condition of gauge invariance.
Using formulae (16) and (18), we list all local terms in the 
specialized sector which arise in the natural splitting of
 the commutator  $\d_\nu^x([T_{1,g}^\nu(x),T_1^g(y)]$ 
according to the procedure described above.:
\begin{eqnarray*} \lefteqn{ \hspace{2,1cm} A \big|_{u\partial_\nu 
\zeta A^\nu \zeta} = (-i) 
g^2 \frac{1}{2} f_{abc} f_{a'b'c} : u_a A_\nu^b 
\zeta_{a'} \partial^\nu \zeta_{b'} : \delta (x-y) } \\
& & \hspace{2,1cm} \mbox{} + (-i) g^2 \frac{1}{4} f_{abc}
 f_{c'a'c} : u_a \zeta_b A_\kappa^{a'} \partial^\kappa \zeta_{c'} : 
\delta (x-y) \hspace{2,7cm} \hbox{(31)}\\
& & \hspace{2,1cm} \mbox{} + (+i) g^2 \frac{1}{4} 
f_{abc} f_{a'b'c} : u_a \zeta_b A_\kappa^{a'} 
 \partial^\kappa \zeta_{b'} : \delta (x-y) \end{eqnarray*}
Using the Jacobi-identity $f_{abc} f_{a'b`c} 
= - f_{ab'c} f_{ba'c} - f_{aa'c} f_{b'bc}$ 
in the first term, one arrives at $\alpha_1 \neq \alpha_2$.
\\
Thus, the St\"uckelberg gauge invariance condition (20) 
already breaks down in second order of perturbation theory 
in tree terms. Note that the constraint of normalizability is essential
for this conclusion. \\
The corresponding breakdown of the 
perturbative unitarity condition (3) can also directly shown.\\
Therefore,  perturbative normalizability and 
physical unitarity cannot be established simultaneously in 
this class of models. But the operator gauge invariance condition 
(2) would be sufficient for physical unitarity in perturbation 
theory (3).\\
At this point a short remark about the relation to the conventional Lagrange
formalism is in order. If one starts with the Stueckelberg
Lagrangean [4,5], the following question naturally arises: Are there 
any point transformation of the fields (which preserve the origin) so that 
this Lagrangean is power counting normalizable in a manifest way.   
It is well-known that two Lagrangean which are related by such a field
transformation have the same S-matrix [16].
From the viewpoint of our analysis we immediately can answer 
this question with no - provided the propagators are not changed - because the asymptotic part of the BRS symmetry
which is only relevant for our analysis is not changed by field transformations
(which preserve the origin). So our argument is in this sense field-coordinate-independent .
\\
 We come back to the Curci-Ferrari model which is 
defined in the causal formalism as the most general
 gauge invariant specific coupling with respect
 to the charge $Q_{CF}$ in (12). Because of the missing 
nilpotency of this charge $Q_{CF}$ these models are not 
expected to be unitary to all orders in perturbation theory.\\
A causal analysis (until second order) of a specific coupling
 which is gauge invariant in respect to the charge 
$Q_{CF}$ in (12)  is given in [17]:
$$ T_1 = \frac{i}{2} gf_{abc} : A_\mu^a A_\nu^b F_c^{\nu\mu} :
- igf_{abc} : A_\mu^a u_b \d^\mu \tilde{u}_c: -  
\frac{i}{2} gf_{abc} : \d_\mu A^\mu_a u_b 
\tilde{u}_c: . \eqno(32)$$
In contrast to the St\"uckelberg model, operator gauge invariance 
can be preserved in second order of perturbation theory  
$$[Q_{CF},T_{n=2}]=div, \eqno(33)$$ but using (10), 
one easily  shows that the perturbative condition 
of physical unitarity (3) in second order
$$ \frac{1}{2} (P_\bot T_2^+(x_1,x_2) + T_2(x_1,x_2) P_\bot) =$$
 $$= P_\bot T_1(x_1) P_\bot T_1(x_2) P_\bot + P_\bot T_1(x_2) 
P_\bot T_1(x_1) P_\bot +div \eqno(34)$$
breaks down  for all $\xi \neq 0$. The case $\xi=0$ needs 
further consideration. The generalizations of these findings 
are straightforward. The general specific coupling which is gauge 
invariant in respect to $Q_{CF}$ in (12) can be written as 
$$T_1 = \frac{i}{2} gf_{abc} : A_\mu^a A_\nu^b F_c^{\nu\mu} :-
 \frac{i}{2} gf_{abc} : A_\mu^a u_b \d^\mu \tilde{u}_c: + $$ $$+\frac{i}{2} 
gf_{abc} : A_\mu^a \d^\mu u_b \tilde{u}_c: + \quad\alpha\quad igf_{abc} 
\d_\mu (: A_a^\mu u_b \tilde{u}_c:), \quad \alpha \quad \mbox{free}. 
\eqno(35)$$
We left out the possible two-operator terms in $T_1$ again. 
The four-operator terms compatible with conditions {\bf (B)-(F)}
 again are ruled out by the gauge invariance condition, {\bf (A)}
   $ [Q_{CF}, T_{n=1}] =div$. But as in the 
massless case [13], the operator gauge invariance condition 
in second order, $\quad [ Q_{CF}, T_{n=2} \big|_{tree,4} ]= div, 
\quad$ uniquely fixes the normalization of $ \quad T_{n=2}
 \big|_{tree,4}$ and naturally introduces a four gluon coupling
 and a four ghost coupling in $T_{n=2}$.
In the perturbative analysis, the most general gauge invariant coupling 
in the general $\xi$-gauge , together with the local normalization terms
 in $T_{n=2} \big|_{tree,4}$, coincide with the interaction terms of
 the Curci-Ferrari Lagrangian - fixed in a linear $\xi$-gauge - which 
is invariant 
 under the full BRS-transformations of the interacting fields and fulfills 
reasonable certain additional symmetry conditions. For example see formula 
(2.1) in [3a].\\
So in contrast to the St\"uckelberg models, the operator gauge 
invariance can be proven to all
orders of perturbation theory, $ [Q_{CF} , T_n] = div$ (2), along the same 
line as in the massless case, but this gauge invariance 
definition with respect to the charge $Q_{CF}$ in (12) does not 
serve as a sufficient 
condition for physical unitarity because of its missing 
nilpotency. However, such models are useful because of 
the well-behaved
$m \rightarrow 0$-limit. As it is proposed in [18], 
such models serve as a good infrared 
regularization of the massless theory. In fact, they also 
constitute  a promising
starting point in the causal approach for the 
investigation of the adiabatic limit $g \rightarrow 1$ 
in the massless theory.  Such an investigation is crucial 
for the analysis of the 
physical infrared problem, which is naturally separated in 
the causal formalism
by adiabatic switching of the $n$-point distributions 
$T_n$ by a tempered testfunction $g$ (see (1)).\\
\\
Summing up, we have presented a short analysis of 
some genuine massive nonabelian 
gauge theories in the Epstein-Glaser approach in order to clarify the
different reasons of the failure of unitarity in these models. Such 
an analysis in the well-defined Fock space of asymptotic fields is
 simplified because the asymptotic (linear) part of the BRS-symmetry
has to be considered only.

{\bf Acknowledgements:} I thank R.Stora for important comments  on
the Epstein-Glaser method, G.Thompson and N.Dragon for useful discussions 
about massive nonabelian gauge theories, G.Scharf for reading the 
manuscript and the Swiss National Science Foundation for 
financial support\\
\vskip1,5cm
{\bf References:}

\begin{tabbing}

1. \quad\quad\quad\=  A. Connes, J. Lott,\\
 \>  Nuclear Physics (Proc. Suppl) B 18 (1990) 29\\
 \>  R. Coquereaux. R. H\"aussling, N. Papadopoulos, F. Scheck,\\
 \>  International Journal of Modern Physics A7 (1992) 2809\\
2. \> R. Delbourgo, S. Twisk, G. Thompson,\\
\> International Journal of Modern Physics A3 (1988) 435\\
3. \>  G. Curci, R. Ferrari\\
 \> Nuovo Cimento 32A (1976) 151, 35A (1976) 1,\\
4.\> T.Kunimasa, T.Goto\\
\> Progress of Theoretical Physics 37 (1967) 452\\ 
\> A. Burnel,\\
\> Physical Review D33 (1986) 2981, D33 (1986) 2985\\
\> and references therein\\
5. \> T. Fukuda, M.Monda, M. Takeda, K. Yokoyama,\\
 \>Progress of Theoretical Physics 66 
(1982) 1827, 67 (1982) 1206, 70 (1983) 284\\
6. \> V. Periwal,\\
\> PUPT-1563, hep-th/9509085\\
7. \> I. Ojima,\\
\> Zeitschrift f\"ur Physik C13 (1982) 173\\
8.\> J. De Boer, K. Skenderis, P. van Nieuwenhuizen, A. Waldron,\\
\> ITP-SB-95-43, hep-th/9510167\\
 9.\>  H. Epstein, V. Glaser,\\
 \>  in G. Velo, A.S. Wightman (eds.):\\
 \>  Renormalization Theory,\\
 \>  D. Reidel Publishing Company, Dordrecht 1976, 193\\
10.\> O. Piguet, A. Rouet,\\
\> Physics Reports 76 (1981) 1\\
\> R. Stora,\\
\> Differential Algebras, ETH-Z\"urich-Lectures (1993), unpublished\\ 
   \> G. Scharf,\\
 \>  Finite Quantum Electrodynamics (Second Edition),\\
 \>  Springer, Berlin 1995\\
11. \> M. D\"utsch, T. Hurth, G. Scharf, \\
 \>  Nuovo Cimento 108A (1995) 679, 108A (1995) 737 \\ 
12. \> T. Hurth,\\
 \>  Annals of Physics 244 (1995) 340, hep-th/9411080\\
13. \> T. Hurth,\\
 \>  ZU-TH-20/95, hep-th/9511139\\
14. \> F. Krahe,\\
 \>  DIAS-STP-95-01, hep-th/9508038\\
15. \> M. D\"utsch, T. Hurth, F. Krahe, G. Scharf,\\
 \>  Nuovo Cimento 106A (1993) 1029\\
16. \>S. Coleman, J. Wess, B. Zumino,\\
 \> Physical Review 177 (1969) 2239\\
17. \> A. Aste, M. D\"utsch, G. Scharf,\\
 \> ZU-TH-27/95\\
18. \>G. Curci, E. d'Emilio,\\
 \>  Physics Letters 83B (1979) 199\\
 \> V. Periwal,\\ 
 \>  PUPT-1562, hep-th/9509084\\
\>  A. Blasi, N. Maggiore,\\
\> UGVA-DPT-95-11-908,hep-th/9511068\\

\end{tabbing}

\end{document}